%% file: main.tex
\documentclass[sigconf]{acmart}

\setcopyright{none}
\renewcommand\footnotetextcopyrightpermission[1]{}

\acmYear{2026}
\copyrightyear{2026}

\usepackage{caption}
\usepackage{graphicx}
\usepackage{tabularx} 
\usepackage[normalem]{ulem}
\usepackage{makecell} 
\usepackage{xspace}
\usepackage{booktabs}
\usepackage{multirow}
\usepackage{cleveref} 
\usepackage{lipsum}
\usepackage{enumitem}
\usepackage{amsmath}
\usepackage{float}
\usepackage{adjustbox}
\setlist[itemize]{noitemsep, topsep=0pt}
\usepackage{adjustbox}
\usepackage{subcaption}
\usepackage[normalem]{ulem}
\usepackage{algorithmicx}
\usepackage{algorithm}     
\usepackage{algpseudocode} 
\usepackage{amsmath}       
\usepackage{booktabs}   
\usepackage{pifont}     
\usepackage{xcolor}     
\usepackage{colortbl}   
\usepackage{graphicx}   

\usepackage{amssymb}

\definecolor{Gray}{gray}{0.9} 

\usepackage{soul}
\usepackage{hyperref}
\hypersetup{
    colorlinks=true,
    linkcolor=blue,
    filecolor=magenta,
    urlcolor=cyan,
}

\begin{document}

\title[]{Now or Never: Continuous Surveillance AIoT System for Ephemeral Events in Intermittent Sensor Networks}


\author{Joonhee Lee}
\email{neo81389@yonsei.ac.kr}
\affiliation{%
\institution{Yonsei University}
\country{Seoul, South Korea}
}

\author{Kichang Lee}
\email{kichang.lee@yonsei.ac.kr}
\affiliation{%
\institution{Yonsei University}
\country{Seoul, South Korea}
}

\author{JeongGil Ko}
\email{jeonggil.ko@yonsei.ac.kr}
\affiliation{%
 \institution{Yonsei University}
 \country{Seoul, South Korea}
}

\begin{abstract}
    \input{sec/0_abstract}
\end{abstract}

\maketitle

\input{sec/1_intro}
\input{sec/2_system}

\input{sec/3_evaluation}

\input{sec/4_conclusion}

\input{sec/5_profile_details}


\bibliographystyle{ACM-Reference-Format}
\bibliography{references, eis-lab}


\end{document}

%% file: sec/0_abstract.tex
Wilderness monitoring tasks, such as poaching surveillance and forest fire detection, require pervasive and high-accuracy sensing. While AIoT offers a promising path, covering vast, inaccessible regions necessitates the massive deployment of maintenance-free, battery-less nodes with limited computational resources. However, these constraints create a critical `Availability Gap.' Conventional intermittent operations prioritize computation throughput, forcing sensors to sleep during energy buffering. Consequently, systems miss ephemeral, `now-or-never' events (e.g., Vocalizations of natural monuments or Fire), which is fatal for detecting rare but high-stakes anomalies. To address this, we propose an Energy-aware Elastic Split Computing Algorithm that prioritizes continuous sensing by dynamically offloading tasks to energy-rich neighbors. Preliminary results demonstrate stable monitoring of an additional $2,496\;\text{m}^2$ and the capture of approximately 103 more critical events per day.
Ultimately, this algorithm establishes a robust foundation for building resilient, fail-safe surveillance systems even on resource-constrained nodes.


%% file: sec/1_intro.tex
\section{Introduction}
\label{sec:Intro}

Wilderness monitoring tasks, such as poaching surveillance and forest fire detection, demand pervasive situational awareness to distinguish critical threats from background noise (e.g., differentiating a poacher's footstep from wind-blown foliage)~\cite{iot_antipoaching, forest_fire_ai}. Simple algorithms and traditional machine learning models often fail to capture complex patterns with sufficient precision. As a result, deploying deep neural networks (DNNs) at the edge has become essential for reliable detection~\cite{baek2025ai, shin2024ictc, sung2024gzip}. However, in these environments, maintaining a stable power supply and performing regular maintenance are challenging. Consequently, sustaining large-scale deployments with battery-powered nodes is logistically infeasible, making energy-harvesting AIoT the only viable solution for pervasive deployment~\cite{eh_wsn_review}.

Unfortunately, such energy-harvesting systems introduce a fundamental challenge to continuous availability, as frequent power interruptions conflict with the computation-intensive nature of DNNs~\cite{enabling_fast_dl}.
Current intermittent computing strategies typically mitigate this tension by accumulating energy and executing tasks only when sufficient energy becomes available, focusing on eventual completion (or forward progress) rather than continuous availability.
We argue that this store-and-compute paradigm may produce latency unsuitable for \emph{ephemeral event detection}.
For instance, a transient gunshot or an initial spark does not persist until a node recharges; in this setting, the latency introduced by energy buffering can effectively constitute mission failure~\cite{mottola_approximate}.

Therefore, the core challenge is to maintain \textbf{continuous spatiotemporal coverage with high precision} despite individual device intermittency.
Unlike prior approaches that overlook \textit{spatial energy variance}, we propose a decentralized algorithm that leverages energy diversity across nodes to enable reliable wide-area surveillance.
By dynamically offloading tasks from energy-starved nodes to energy-rich neighbors, our approach sustains continuous surveillance while preserving high precision.
The main contributions are summarized as follows:

\begin{itemize}[leftmargin=*]
    \item \textbf{Coverage-First Scheduling for Continuous Surveillance.}
    We propose a novel intermittent-computing scheduling algorithm that minimizes sensing blind spots via lightweight state propagation, avoiding energy-expensive bidirectional handshakes. Our ``Coverage First'' scheduler coordinates wake-up cycles so that at least one node remains active in each target region, enabling instantaneous capture of ephemeral events even under unstable energy-harvesting conditions.
    
    \item \textbf{Energy-Aware Collaborative Inference.}
    We introduce a collaborative computing algorithm for DNN inference on power-constrained wireless sensor networks, driven by \textit{lightweight neighbor energy prediction}. The predicted energy states guide dynamic task partitioning and offloading, enabling effective orchestration across nodes to bypass single-node limitations.
    
    \item \textbf{Hardware-Profiled High-Fidelity Simulation.}
    We validate the feasibility of our algorithm through a rigorous simulation grounded in real-world hardware profiling. We conducted extensive power profiling of the Micro Controller Unit (MCU) across all operational states—including sensing, computation, transmission, and idle modes. By modeling a network of 100 nodes with precise energy harvesting and discharging dynamics, we show the feasibility of our decentralized approach in energy-constrained environments.
\end{itemize}

%% file: sec/2_system.tex
\section{System}
\label{sec:System}
\begin{figure}[t]
    \centering
    \includegraphics[width=0.95\linewidth]{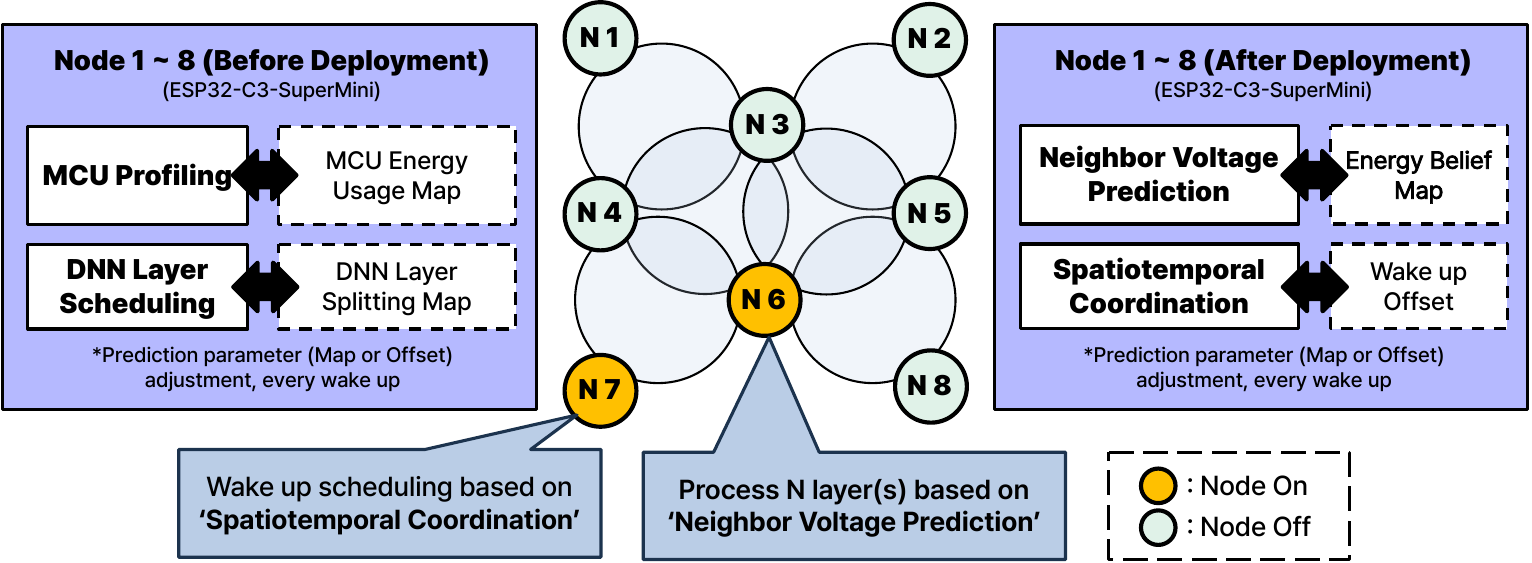} 
    \caption{Operational Algorithm for Energy-Aware Collaborative Inference}
    \label{fig:system_overview}
\end{figure}

We propose a decentralized, continuous surveillance coordination algorithm designed specifically for intermittent sensor networks operating under heterogeneous harvesting conditions.
As illustrated in Figure~\ref{fig:system_overview}, each nodes learn heterogeneous environment of neighbor node and adapts its life-cycle that dynamically adjusts sensing duration, computation depth, and wake-up schedules based on energy availability and neighbor states.

\subsection{Pre-processing Before Deployment}
Prior to deployment, the system undergoes an offline profiling phase.
We operate under the assumption that all nodes within the network possess homogeneous hardware specifications, utilizing identical sensors and MCUs.
This homogeneity underpins the Neighbor Voltage Prediction mechanism detailed in the post-deployment phase of Figure~\ref{fig:system_overview}.
By utilizing an \textit{MCU Energy Usage Map} generated through offline profiling, the system operates on the premise that nodes performing similar computational tasks under comparable environmental conditions will exhibit correlated energy consumption patterns.

Furthermore, we construct a static \textit{DNN Layer Splitting Map} via offline scheduling.
Based on the profiled energy cost for specific computational loads, the target DNN model is pre-partitioned into logical layer groups.
During runtime inference, nodes utilize the Energy Usage Map to estimate both their own and their neighbors' energy states.
Guided by the Splitting Map, the system adaptively determines the optimal partition point to ensure network sustainability. 
Specifically, if a node determines it possesses a higher energy surplus than its neighbors, it locally executes a larger portion of the DNN layers before offloading the intermediate results to adjacent nodes.

\subsection{Energy-Adaptive Life-Cycle with Hysteresis}
To prevent the ``zombie node'' phenomenon where nodes repeatedly boot and fail due to insufficient energy, we implement a hysteresis-based recovery mechanism. 
By leveraging profiled MCU energy consumption, this mechanism estimates the required energy buffer for failure-free operation and the maximum sustainable active duration. Specifically, the node enters \textit{deep sleep} when its energy level drops below a critical lower threshold (e.g., 1\%) and remains inactive until it recharges to a sufficient recovery threshold (e.g., 20\%).


Once active, the node employs an adaptive sensing strategy. Instead of a fixed sensing window, the node calculates the maximum possible sensing duration ($T_{sense}$) allowed by its residual energy, after reserving a safety margin for essential computation and transmission tasks.
This approach maximizes the system's total surveillance coverage during energy-rich periods.

\subsection{Neighbor Voltage Prediction (NVP)}
To facilitate cooperation without energy-intensive handshakes, our system utilizes a piggybacking-like mechanism. Every communication packet contains a small header field representing the sender's current voltage ($V_{sender}(t)$) at time $t$. This allows neighbors to build a local ``Energy-Belief Map'' without dedicated synchronization packets. 

Crucially, we employ a Neighbor Voltage Prediction (NVP), which is a self-referenced prediction. This approach leverages the spatial consistency of environmental energy; because adjacent nodes typically share similar solar irradiance angles and harvesting conditions, a node $i$ predicts the current voltage of a neighbor $j$, denoted as $\hat{V}_j(t)$, based on its own voltage change $\Delta V_i$ from $t_{last}$ to $t$:
\begin{equation}
    \hat{V}_j(t) = V_j(t_{last}) + \Delta V_i(t) + \delta_{offset}
\end{equation}
where $V_j(t_{last})$ is the last known voltage of neighbor $j$, and $\delta_{offset}$ is a learned correction term. This allows nodes to accurately estimate neighbor availability and select the optimal offloading partner.

\subsection{Decentralized Spatiotemporal Coordination}
Traditional collaborative sensing often results in overlapping detection zones that waste redundant energy. Our algorithm addresses this through two decentralized mechanisms:

\begin{itemize}[leftmargin=*]
    \item \textbf{Collision-Aware Desynchronization:} If a node receives a signal from a neighbor \textit{during} its sensing phase, it identifies a spatiotemporal overlap. To resolve this, the node shifts its next wake-up schedule by applying a random backoff time, effectively separating the sensing windows of adjacent nodes.
    \item \textbf{Collaborative Offloading:} During the handover window (immediately after sensing), if a node receives a task but lacks sufficient energy to complete DNN inference, it processes only the feasible initial layers. The intermediate feature tensors are then offloaded to the neighbor with the highest predicted energy surplus ($\hat{V}_{neighbor} > V_{self}$), ensuring tasks naturally flow towards energy-rich nodes.
\end{itemize}

%% file: sec/3_evaluation.tex
\section{Evaluation}
\label{sec:Evaluation}

\subsection{Simulation Setup}
We developed a Python-based simulator to evaluate the energy efficiency and feasibility of the proposed system. The simulation creates virtual nodes modeled after the \textbf{ESP32-C3-SuperMini}, a microcontroller unit (MCU) featuring a 32-bit RISC-V CPU clocked at 160 MHz and 400 KB of SRAM. It experimentally tracks how data is processed, partitioned, and propagated through the network from input to output.

\begin{figure}[!t]
\centering
\subcaptionbox{Simulated Node Map}[0.59\linewidth]{
    \includegraphics[width=1.0\linewidth]{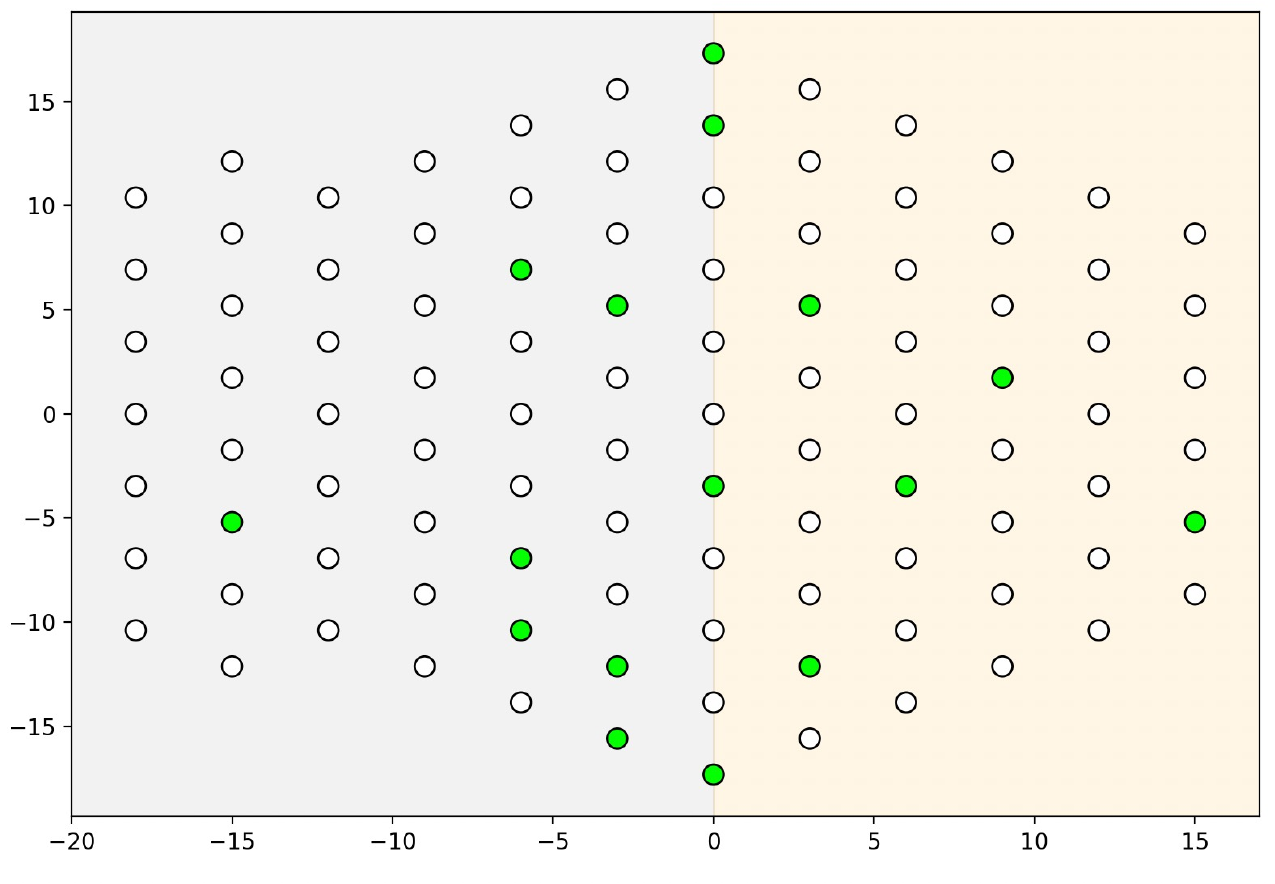} 
}
\hfill
\subcaptionbox{MCU Testbed}[0.39\linewidth]{
    \raisebox{34.5mm}{
        \includegraphics[width=0.97\linewidth, angle=-90]{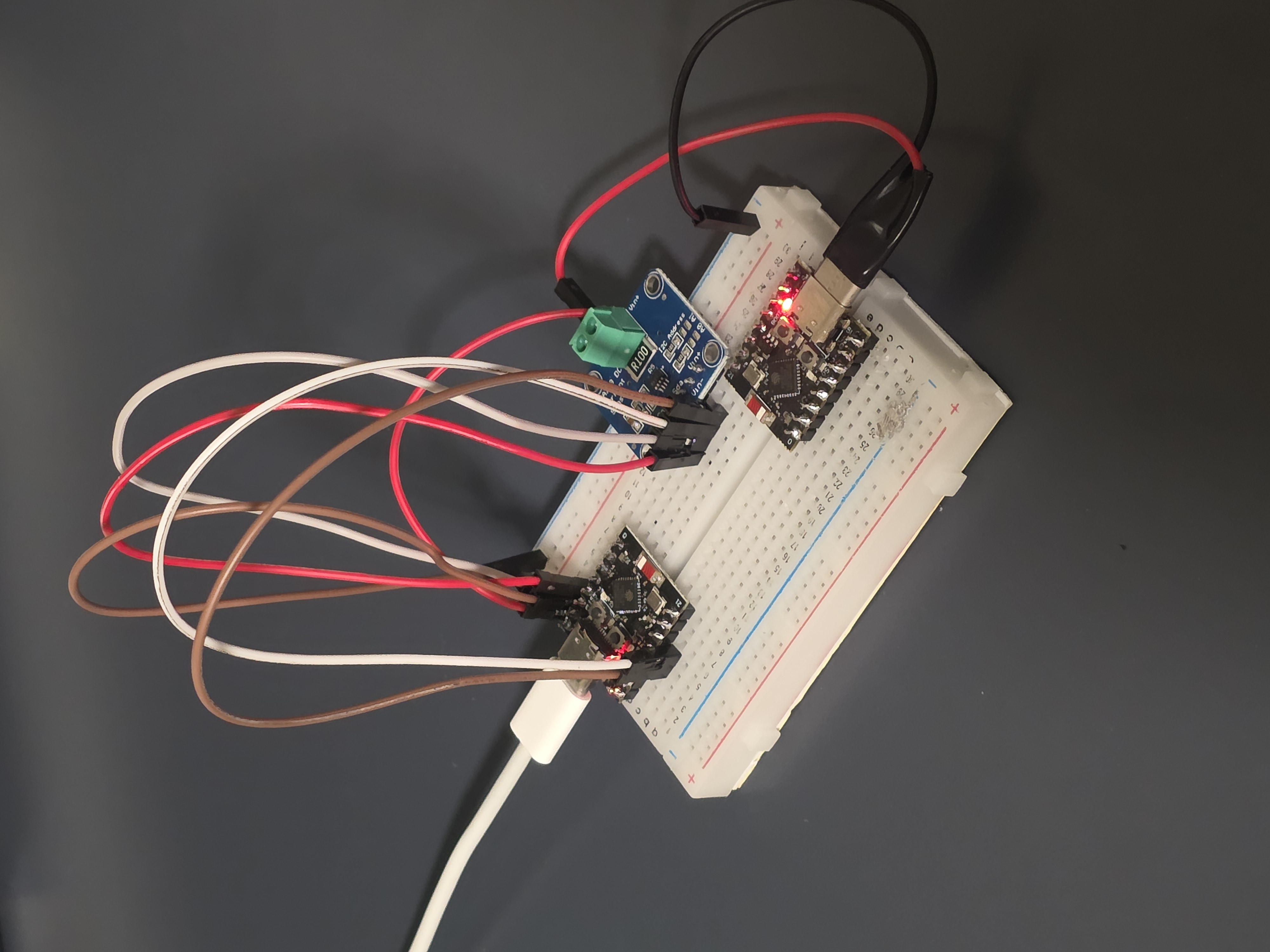}
    }
}
\caption{Simulation and Profiling Setup}
\vspace{-2ex}
\label{fig:simulation_hw_setup}
\end{figure}

We assumed a network topology where all nodes are arranged in a hexagonal grid (honeycomb structure) with uniform spacing.
We employ \texttt{ESP-NOW} as a communication protocol that is widely used in IoT and wireless sensor network deployment.
While the manufacturer states a maximum line-of-sight range of 200 meters for ESP-NOW~\cite{espnow_reliability}, we conservatively set the maximum effective communication range to 20 meters, considering the antenna performance of the SuperMini board and potential real-world deployment constraints. 
Consequently, the simulation assumes a uniform inter-node distance of 20 meters.
Within this coverage area, we assume that 50 random acoustic events occur uniformly per hour across the regions formed by the nodes. The simulation is conducted for 10,000 seconds ($\approx$ 2.78 hours), during which approximately 138 acoustic events occur. To simulate environmental heterogeneity, 100 nodes are randomly distributed between sunny and shady conditions in a 50:50 ratio. The battery capacity is fixed at 50 mAh, with solar harvesting rates set to 300 $\mu$W for sunny areas and 50 $\mu$W for shady areas.

For DNN inference, we employ a five-layer CNN tailored to audio-based event detection.
To support distributed execution, the network is designed to be partitioned at any layer boundary.
We evaluate on the ESC-10 dataset using a ten-class classification setting, where each sample is a 5-second audio recording of environmental sounds, including animal and insect vocalizations.

In the baseline \textit{Vanilla Mode}, a node processes only a single fixed layer before transmitting the intermediate tensor to the next node.
In contrast, our proposed \textit{Algorithm Mode} enables dynamic layer processing, where a node adaptively determines the number of layers to compute locally based on its battery status.


To enable realistic energy simulations, we performed extensive power profiling of the MCU across idle, sensing, communication, and computation modes. We measured a baseline power draw of 434.85~mW when the device is active but idle. For data transmission, the energy cost depends on the payload size, ranging from 61.39 to 287.80~$\mu$J/Byte. For computation, a full forward pass of the five-layer DNN consumes 6.07~mJ. During sensing, the device draws 439.48 to 505.95~mW continuously. Also, we model sensing power using the average power draw of representative sensors compatible with this MCU.
For completeness, we provide a detailed breakdown of the profiling results by operation mode in Section~\ref{sec:profile_details}.

\subsection{AIoT Network Performance and Impact on Quality Surveillance}

In an intermittent computing environment, if sensor nodes wake up arbitrarily without neighbor awareness, spatial optimization becomes impossible, leading to redundant sensing in the same area while leaving other areas unmonitored. Furthermore, ignoring the heterogeneous energy harvesting conditions (e.g., solar panel placement) exacerbates system instability. To address this, we propose a decentralized coordination scheme governed by two key mechanisms: a scheduling adjustment rule for desynchronization (Eq. \ref{eq:desync}) and an energy-aware offloading criterion (Eq. \ref{eq:offloading}).

First, to minimize spatial sensing redundancy, a node must detect if its active window overlaps with a neighbor. If a node receives a beacon/signal from neighbor $j$ during its sensing phase $t_{sense}$, it implies a collision in the wake-up schedule. It then adjusts the next sleep interval $T_{sleep}$ as follows:

\begin{equation}
    T_{sleep}^{(k+1)} = T_{base} + I(overlap) \cdot \Delta_{backoff}
    \label{eq:desync}
\end{equation}

where $T_{base}$ is the default cycle time, $I(\cdot)$ is the indicator function which is 1 if an overlap is detected and 0 otherwise, and $\Delta_{backoff}$ is a random time shift derived from a uniform distribution $U(t_{min}, t_{max})$. This stochastic shift effectively desynchronizes nodes covering the same area over time.

Second, to optimize computational distribution via energy availability, we introduce a decision criterion for task offloading. When a node $i$ receives a task request along with the sender's voltage information $V_j$, it compares its residual voltage $V_i$ to decide the extent of local processing. The decision to accept and process $L$ layers locally is governed by:

\begin{equation}
    \text{Process Locally if } \quad (V_i - V_{req}(L)) > (V_j + \alpha)
    \label{eq:offloading}
\end{equation}

where $V_{req}(L)$ is the estimated voltage drop required to process $L$ layers, and $\alpha$ is a hysteresis threshold to prevent oscillation. This ensures that tasks naturally flow from energy-poor nodes to energy-rich nodes.

\begin{figure}[!t]
\centering
\subcaptionbox{Coverage and Stability Performance}[0.49\linewidth]{
    \includegraphics[width=1.0\linewidth]{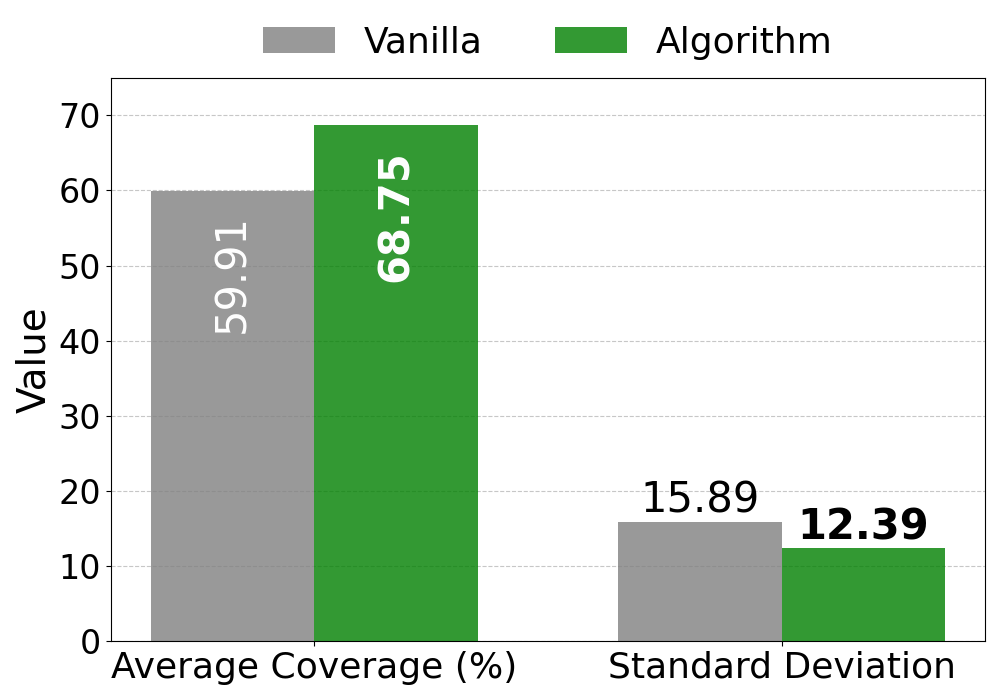} 
}
\hfill
\subcaptionbox{Environment Impact of Sensing}[0.49\linewidth]{
    \includegraphics[width=1.0\linewidth]{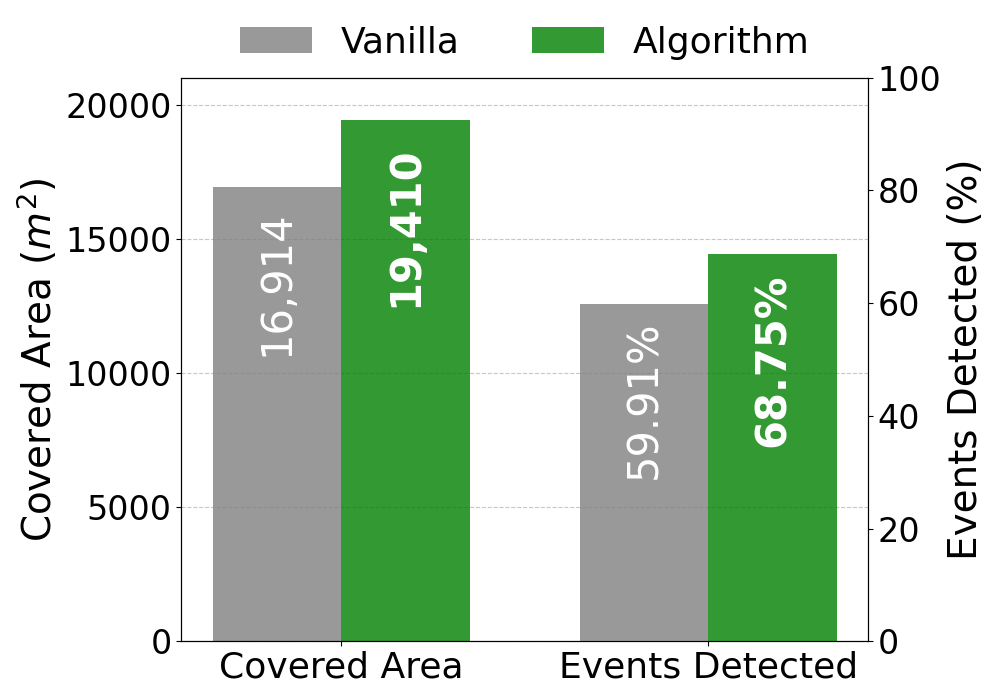}
}
\vspace{-2ex}
\caption{Simulation results showing coverage stability and environmental impact.}
\vspace{-2ex}
\label{fig:simulation_result}
\end{figure}

Figure \ref{fig:simulation_result}(a) compares the performance of the \textit{Vanilla Mode} against our proposed \textit{Algorithm Mode}. In the Vanilla Mode, a node wakes up only when its battery is sufficiently charged to support at least 3 seconds of sensing, the first layer of inference, and data transmission. It remains active until its energy depletes to the minimum transmission threshold, broadcasts its data, and then enters deep sleep. This approach is purely reactive to individual battery states.

In contrast, the \textit{Algorithm Mode} dynamically adapts both the wake-up schedule and the computational load (i.e., number of layers processed) based on Equations \ref{eq:desync} and \ref{eq:offloading}. By considering environmental energy disparities and spatial overlaps, the proposed algorithm achieved a 9.84\% increase in average coverage (from 59.51\% to 68.75\%). Furthermore, the standard deviation of coverage dropped from 15.89 to 12.39, indicating a much more stable and reliable surveillance performance. These results demonstrate that a simple decentralized algorithm can effectively mitigate spatial inefficiencies caused by uncoordinated intermittent computing and heterogeneous energy supplies.

Figure ~\ref{fig:simulation_result}(b) translates these performance gains into physical metrics, assuming the realistic communication range of 20 meters for the ESP32-C3-SuperMini. The simulation results indicate that the proposed system effectively covers an additional average area of $2,496 m^2$ compared to the baseline. In the context of our simulation scenario, this increased coverage implies the ability to capture approximately 9\% more events on average that would have otherwise gone undetected due to nodes being offline or redundant sensing.

\subsection{Performance Analysis of Split DNN Inference}

\begin{figure}[t]
    \centering
    \includegraphics[width=0.90\linewidth]{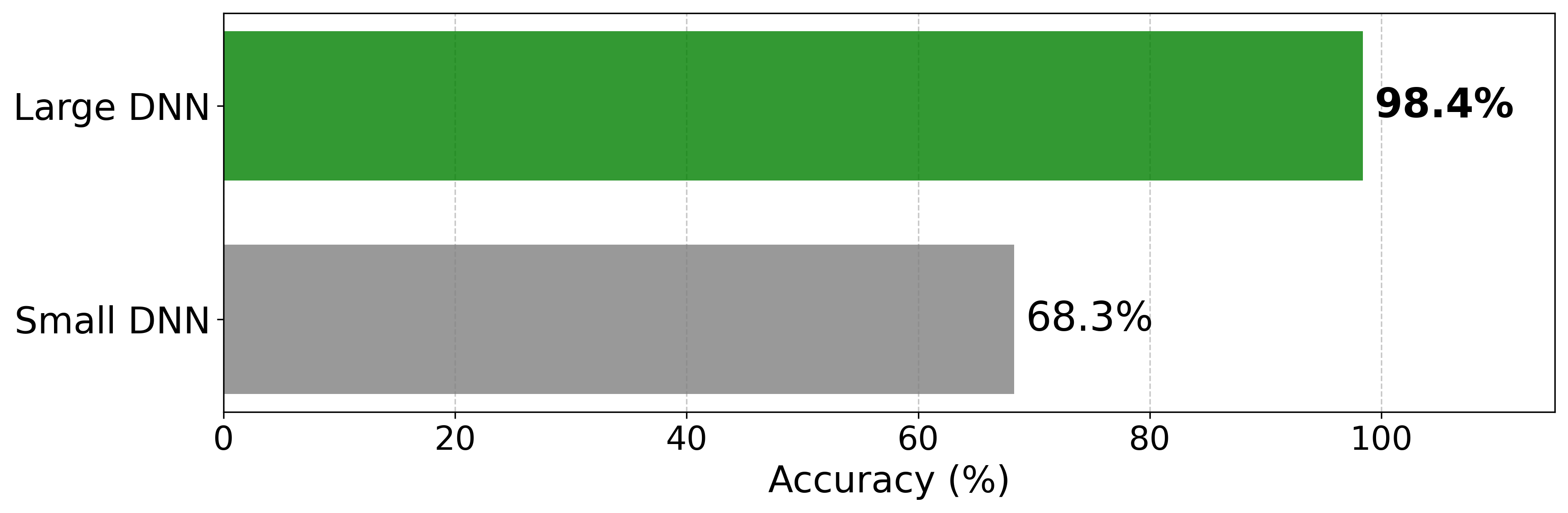} 
\vspace{-3ex}
    \caption{Accuracy comparison for varying model size.}
    \label{fig:audio_detection_performance}
\vspace{-3ex}
\end{figure}

The experiment presented in Figure \ref{fig:audio_detection_performance} evaluates the performance trade-offs between a lightweight model, capable of standalone operation on a resource-constrained node like the ESP32-C3-SuperMini, and a larger model where inference is collaboratively distributed across multiple nodes.

The comparison baseline, referred to as the \textit{Small DNN}, processes spectrogram images resized to $32 \times 32$. Its architecture consists of two convolutional (Conv2D) layers with an intermediate Global Average Pooling layer. When compiled with its parameters, this model occupies approximately 75 KB of memory, making it compact enough to be fully deployed on a single ESP32-C3-SuperMini unit.

Experimental results indicate that the standalone Small DNN achieved an accuracy of 68.3\%. In contrast, the distributed scenario, which executes the larger model by splitting layers across nodes, demonstrated a significantly superior accuracy of 98.4\%. This substantial performance gap validates the efficacy and necessity of the proposed split-computing architecture for high-fidelity inference in resource-constrained networks.

%% file: sec/4_conclusion.tex
\section{Conclusion and Future Research}
\label{sec:Conclusion}

In this work, we addressed the intelligence gap and intermittent sensing issues in battery-less AIoT by proposing a decentralized, energy-aware elastic split computing algorithm. Unlike traditional intermittent systems that view "waiting for energy" as inevitable, our approach prioritizes continuous surveillance to capture ephemeral events in the wild. Preliminary results demonstrate that our cooperative timing-first algorithm significantly enhances surveillance stability, covering an additional $2,496 \text{ m}^2$ and capturing approximately $103$ more time-critical events per day compared to vanilla methods.

As an ongoing research effort, this algorithm aims to realize a truly ubiquitous AIoT system where high-performance DNN models are seamlessly executed across distributed environments. By dynamically partitioning computation layers according to spatial energy variance, we showed promising result of resolving the bottleneck of individual node failures, ensuring that sophisticated DNN inference is possible even under heterogeneous harvesting conditions. However, our evaluation indicates that the system has not yet reached its theoretical optimum. As the current coordination algorithm relies on heuristic approaches to prioritize feasibility, there remains significant potential for further algorithmic refinement to maximize performance.

%% file: sec/5_profile_details.tex
\section{Profile Details}
\label{sec:profile_details}

This profile details present the power and energy profiling data used in the simulation for the ESP32-C3-SuperMini. The energy consumption was measured and modeled to accurately reflect the constraints of batteryless operation.

\subsection{Mode 0: Idle State}
\label{sec:idle_profile}
The idle state represents the baseline power consumption when the MCU is active but not performing inference or transmission.
\begin{itemize}
    \item \textbf{Average Power Consumption:} $434.85 \text{ mW}$
\end{itemize}

\subsection{Mode 1-5: Layer-wise Inference Profiling}
\label{sec:inference_profile}
Table \ref{tab:layer_profile} summarizes the computational cost, latency, and energy consumption for processing each layer of the ResNet-based Audio Detection model.

\begin{table}[h]
    \centering
    \caption{Layer-wise Inference Performance and Energy Consumption}
    \label{tab:layer_profile}
    \resizebox{\linewidth}{!}{%
    \begin{tabular}{lcccccc}
        \toprule
        \textbf{Layer} & \textbf{FLOPs} & \textbf{Latency} & \textbf{Total Energy} & \textbf{Avg Power} & \textbf{Output Size} & \textbf{Efficiency} \\
         & & (ms) & ($\mu J$) & (mW) & (bytes) & (mJ/MFLOP) \\
        \midrule
        Layer 0 & 232,800 & 1,305 & 577,770 & 459.96 & 400 & 2,481.82 \\
        Layer 1 & 291,000 & 1,603 & 800,620 & 499.29 & 320 & 2,751.27 \\
        Layer 2 & 436,500 & 2,400 & 1,197,630 & 498.79 & 240 & 2,743.70 \\
        Layer 3 & 582,000 & 3,207 & 1,589,970 & 495.50 & 160 & 2,731.91 \\
        Layer 4 & 727,500 & 3,845 & 1,901,280 & 494.75 & 80  & \textbf{2,613.45} \\
        \bottomrule
    \end{tabular}%
    }
\end{table}

\subsection{Mode 6: Transmission Profiling}
\label{sec:tx_profile}
Table \ref{tab:tx_profile} details the energy cost required to transmit the output tensor of each layer via ESP-NOW.


\begin{table}[h]
    \centering
    \caption{Transmission Energy Cost per Layer Output}
    \label{tab:tx_profile}
    \begin{adjustbox}{width=.9\linewidth}
    \begin{tabular}{lccccc}
        \toprule
        \textbf{Source} & \textbf{Payload} & \textbf{Time} & \textbf{Power} & \textbf{Total Energy} & \textbf{Energy Density} \\
        \textbf{Layer} & (bytes) & (ms) & (mW) & ($\mu J$) & ($\mu J$/Byte) \\
        \midrule
        Layer 0 & 400 & 30.50 & 838.25 & 24,555 & 61.39 \\
        Layer 1 & 320 & 138.00 & 731.17 & 70,284 & 219.64 \\
        Layer 2 & 240 & 137.61 & 825.40 & 69,073 & 287.80 \\
        Layer 3 & 160 & 26.75 & 887.37 & 23,596 & 147.48 \\
        Layer 4 & 80  & 27.80 & 806.26 & 22,226 & 277.83 \\
        \bottomrule
    \end{tabular}
    \end{adjustbox}
\end{table}

\subsection{Mode 7: Sensor Energy Profiling}
\label{sec:sensor_profile}
We profiled the continuous energy consumption of various sensor configurations over a 3-second active window.
\begin{table}[h]
    \centering
    \caption{Sensor Power Consumption (3-second Active Window)}
    \label{tab:sensor_profile}
    \begin{adjustbox}{width=0.8\linewidth}
    \begin{tabular}{lcc}
        \toprule
        \textbf{Sensor Configuration} & \textbf{Avg Power} & \textbf{Total Energy (3s)} \\
         & (mW) & ($\mu J$) \\
        \midrule
        Ultrasonic & 449.66 & 1,348,970 \\
        Accel + Gyro + Mag & 505.95 & 1,517,850 \\
        \bottomrule
    \end{tabular}
    \end{adjustbox}
\end{table}
